\documentclass[11pt,a4paper]{article}
\usepackage{jheppub}
\usepackage{color,subfigure}

\title{Light mesons around critical end points in $T-\mu_B-\mu_I-eB$ space}
\author{Luyang Li$^1$,
        Shijun Mao$^2$%\footnote{Corresponding author}
        }
\affiliation[]{$^1$ Xi'an University of Posts and Telecommunications, Xi'an, Shaanxi 710121 China\\
$^2$ Institute of Theoretical Physics, School of Physics, Xi'an Jiaotong University, Xi'an, Shaanxi 710049, China}
\emailAdd{maoshijun@mail.xjtu.edu.cn}

\abstract{%Chiral symmetry restoration and pion superfluid phase transition are investigated in magnetized NJL model. Different from the conventional study of order parameters, we investigate the meson properties in $T-\mu_B-\mu_I-eB$ space, paying special attention to the (pseudo-)Goldstone mode.
Light mesons $(\sigma, \pi^0, \pi^\pm)$ are investigated in $T-\mu_B-\mu_I-eB$ space by using a two-flavor NJL model, which are related to the chiral symmetry restoration and pion superfluid phase transition. In $T-\mu_B-eB$ space, during the chiral restoration process, the mass of pseudo-Goldstone mode $\pi^0$ keeps increasing, together with the sudden mass jump. At the critical end point region, $\pi^0$ meson has a very sharp but continuous mass increase, together with a sudden mass jump at the Mott transition, and in the first order chiral phase transition region nearby, we observe twice $\pi^0$ mass jumps, induced by the Mott transition and quark mass jump, respectively. The mass of Higgs mode $\sigma$ first decreases and then increases associated with the chiral symmetry restoration, and shows a jump at the first order chiral phase transition. We plot the chiral phase diagram in terms of the change of quark mass, the Mott transition of pseudo-Goldstone mode $\pi^0$ and the minimum mass of Higgs mode $\sigma$. Due to the explicit breaking of chiral symmetry in physical case, the chiral restoration phase boundaries in $T-\mu_B$ plane from the order parameter side and from the meson side are different from each other. In $T-\mu_I$ plane, the competition between pion superfluid phase transition and chiral symmetry restoration under magnetic fields is studied in terms of the Goldstone mode $\pi^+$ and the pseudo-Goldstone mode $\pi^0$. The isospin symmetry is strict, and the pion superfluid phase transition is uniquely determined by the massless Goldstone mode $\pi^+$. The separation of the two phase boundaries is enhanced by the external magnetic field. Different from the twice mass jumps of $\pi^0$ in the first order chiral phase transition region, the $\pi^+$ meson displays several mass jumps in the chiral crossover region. At the critical end point, $\pi^+$ also shows very sharp but continuous mass changes, together with a mass jump at the Mott transition. Such mass jumps may result in some interesting phenomena in relativistic heavy ion collisions, for instance, the enhancement of pions. And how to extract the experimental signal for the critical end point will be investigated in the future.}

\begin{document}
\maketitle

\section{Introduction}
\label{intro}
The phase structure of Quantum Chromodynamics (QCD) at finite magnetic field, temperature and density attracts much attention in recent years, due to its relation to the evolution of the early Universe, Relativistic Heavy Ion Collisions and compact stars~\cite{beffect,beffect1,beffect2}. The mechanism for a continuous phase transition is the spontaneous symmetry breaking. One can define an order parameter which changes from nonzero value to zero or vice verse when the phase transition happens. On the other hand, the spontaneous breaking of a global symmetry manifests itself in the Goldstone's theorem~\cite{gold1,gold2}: Whenever a global symmetry is spontaneously broken, massless fields, known as Goldstone bosons, emerge. The modifications of hadron properties in medium will also help to understand the QCD phase transitions. %Since neutral (charged) pions are Goldstone bosons of the chiral (isospin) symmetry breaking, the modifications of their properties under external magnetic field will help to understand the chiral restoration (pion superfluid) phase transition. They are extensively investigated, at hadron level~\cite{c1,c3,hadron1,hadron2,qm1,qm2,sigma1,sigma2,sigma3,sigma4,sigma5}, such as with chiral perturbation theory and linear sigma model, and at quark level~\cite{l1,l2,l3,l4,lqcd5,lqcd6,ding2008.00493,njl2,meson,mfir,ritus5,ritus6,mao1,mao11,mao2,wang,coppola,phi,liuhao3,he,maocharge,maopion,yulang2010.05716,q1,q2,q3,q4,huangamm1,huangamm2,q5,q6,q7,q8,q9,q10}, such as in lattice QCD simulations and Nambu--Jona-Lasinio (NJL) model.
%Notable examples for systems in nature that exhibit strong magnetic fields include off-central heavy-ion collisions, the inner core of magnetized neutron stars and, possibly, the early stage of the evolution of our universe~\cite{}.

The electromagnetic interaction provides a sensitive probe of the hadron structure. At hadron level without considering the inner structure, neutral hadrons are blind to electromagnetic fields and their properties in hot and dense medium are not affected by the fields. At quark level, however, the electromagnetic interaction of the charged constituent quarks leads to a sensitive dependence of the neutral meson properties on external electromagnetic fields~\cite{l1,l2,l3,l4,lqcd6,ding2008.00493,njl2,meson,mfir,ritus5,ritus6,mao1,mao11,mao2,wang,coppola,phi,liuhao3,he,maocharge,maopion,yulang2010.05716,q1,q2,q3,q4,huangamm1,huangamm2,q5,q6,q7,q8,q9,q10}. Having the magnetic field of strength compatible with the strong interaction, such as $eB\sim m^2_\pi$, the quark structure of hadrons should be taken into account. In this paper, we will investigate the light meson ($\sigma, \pi^0, \pi^\pm$) properties in $T-\mu_B-\mu_I$ space under external magnetic fields in frame of the two-flavor NJL model at quark level. Within this model~\cite{njl2,njl1,njl3,njl4,njl5}, quarks are treated in mean field level and mesons are the quantum fluctuations constructed from the quark bubble. One can study the QCD phase transitions at the order parameter level~\cite{njl2,njl1,njl3,njl4,njl5,zhuang,model9,model91,model14}. Moreover, the associated hadronic mass spectrum and the static properties of mesons are described remarkablely well~\cite{njl2,njl1,njl3,njl4,njl5,zhuang,model9,model91,model14}.% $\pi^0$ and $\sigma$ are the pseudo-Goldstone mode and Higgs mode of chiral symmetry breaking, respectively, and the charged pion is the Goldstone mode of isospin symmetry breaking.

In temperature-baryon chemical potential $(T-\mu_B)$ plane, chiral symmetry restoration is a smooth crossover in high $T$ and low $\mu_B$ region, and a first order phase transition in low $T$ and high $\mu_B$ region, which are connected by a critical end point~\cite{beffect,beffect1,beffect2,zhuang,mao11,cep1,cep2}. The pseudo-Goldstone mode of chiral symmetry breaking is the neutral pion. During the chiral crossover at vanishing $\mu_B$, the pseudo-Goldstone mode $\pi^0$ shows a mass jump at the Mott transition, because of the discrete Landau level of constituent quarks~\cite{mao2,q2,q3,yulang2010.05716}. With the first order chiral phase transition at vanishing $T$, the pseudo-Goldstone mode $\pi^0$ shows a mass jump at the phase transition point, induced by the corresponding mass jump of constituent quarks~\cite{mao2}. The meson properties around the critical end point is still not clear to us. In temperature-isospin chemical potential $(T-\mu_I)$ plane, there exists the competition between chiral symmetry restoration and pion superfluid phase transition~\cite{maopion,model9,model91,model14,lqcd1,lqcd11,lqcd12,lqcd2,lqcd3,model1,model101,model2,model3,model4,model5,model6,model7,model8,model10,model11,model12,model13,model15,model16,model17,model18,model19,model21,model22,model23,magpion1,magpion2,magpion3}. The Goldstone mode of isospin symmetry breaking is the charged pion. When pion superfluid phase transition happens at $\mu_I>0$, the $\pi^+$ meson becomes massless~\cite{maopion,model9,model91,model14,lqcd2,model12,model13,model21}. Due to the discrete Landau level and different electric charges of constituent quarks, charged pions show several mass jumps as a function of temperature with vanishing $\mu_I$~\cite{maocharge}. Much less is known about the properties of charged pions in $T-\mu_I$ plane.

After the introduction, we generalize the NJL framework of quarks and mesons to $T-\mu_B-\mu_I-eB$ space in Sec.\ref{sec:f}. Numerical results and physical discussions are written in Sec.\ref{sec:rd}, where we focus on the phase diagram and the corresponding (pseudo-)Goldstone mode and Higgs mode. And finally, we come to a brief summary in Sec.\ref{sec:sum}.

\section{NJL Framework}
\label{sec:f}
The two-flavor NJL model is defined through the Lagrangian density in terms of quark fields $\psi$~\cite{njl2,njl1,njl3,njl4,njl5}
\begin{equation}
\label{njl}
{\cal L} = \bar{\psi}\left(i\gamma_\mu D^\mu-m_0+\gamma_0 \mu\right)\psi+G \left[\left(\bar\psi\psi\right)^2+\left(\bar\psi i\gamma_5{\vec \tau}\psi\right)^2\right],
\end{equation}
where the covariant derivative $D_\mu=\partial_\mu+iQ A_\mu$ couples quarks with electric charge $Q=diag (Q_u,Q_d)=diag (2e/3,-e/3)$ to external magnetic field in $z$ direction ${\bf B}=(0, 0, B)$ through the potential $A_\mu=(0,0,Bx_1,0)$, $m_0$ is the current quark mass characterizing the explicit chiral symmetry breaking, the quark chemical potential $\mu
=diag\left(\mu_u,\mu_d\right)=diag\left(\mu_B/3+\mu_I/2,\mu_B/3-\mu_I/2\right)$ is a matrix in the flavor space, with $\mu_u$ and $\mu_d$
being the $u$- and $d$-quark chemical potentials and $\mu_B$ and
$\mu_I$ being the baryon and isospin chemical potentials, and $G$ is the coupling constant in scalar and pseudo-scalar channels. Under external magnetic field, the $SU(2)_L \otimes SU(2)_R$ chiral symmetry is explicitly reduced to $U(1)_L \otimes U(1)_R$. The order parameter of spontaneous chiral symmetry breaking is the neutral chiral condensate $\langle\bar\psi\psi\rangle$ or the (dynamical) quark mass $m_q=m_0-2G\langle\bar\psi\psi\rangle$, and the corresponding Goldstone boson is the neutral pion $\pi^0$. At finite isospin chemical potential $(\mu_I>0)$, with the spontaneous breaking of the isospin symmetry $U(1)_I$ by the charged pion condensate $\langle\bar\psi\gamma_5 \tau^1 \psi\rangle$, the Goldstone boson is $\pi^+$ meson.

There are two equivalent ways to treat the particle propagators under external magnetic field, the Ritus scheme~\cite{ritus1,ritus2,ritus21} and the Schwinger scheme~\cite{beffect,beffect1,beffect2}. For convenience to study both neutral and charged particles, we make derivations with Ritus Scheme in the following, where one can well-define the Fourier-like transformation for the particle propagator between the conserved Ritus momentum space and the coordinate space~\cite{ritus1,ritus2,ritus21}. For example, the quark propagator $S_f(x,y) $ with flavor $f$ in coordinate space can be written as
\begin{eqnarray}
\label{quark}
S_f(x,y) &=& \sum_n \int {d^3\tilde p\over (2\pi)^3} e^{-i \tilde p\cdot (x-y)} P_n(x_1,p_2)D_f(\bar p) P_n(y_1,p_2),\\
P_n({x_1},p_2) &=& {1\over 2}\left[g_n^{s_f}({x_1},p_2)+I_n g_{n-1}^{s_f}({x_1},p_2)\right]+{is_f\over 2}\left[g_n^{s_f}({x_1},p_2)- I_n g_{n-1}^{s_f}({x_1},p_2)\right]\gamma_1 \gamma_2,\nonumber\\
D_f^{-1}(\bar p) &=& \gamma \cdot \bar p-m_q,\nonumber
\label{ritusp}
\end{eqnarray}
where $\tilde p=(p_0,0,p_2,p_3)$ is the Fourier transformed momentum, $\bar p=(p_0,0,-s_f \sqrt{2n|Q_f B|},p_3)$ is the conserved Ritus momentum with $n$ describing the quark Landau level in magnetic fields, $D_f(\bar p)$ is the quark propagator in Ritus momentum space, %$s_f=\text{sgn}(Q_f B)$
$s_f=\text{{sgn}}(Q_f B)$ is the quark sign factor,  $I_n=1-\delta_{n0}$ is governed by the Landau energy level, and the magnetic field dependent function $g_n^{s_f}({x_1},p_2) = \phi_n(x_1-s_f p_2 /|Q_f B|)$ is controlled by the Hermite polynomial $H_n(\zeta)$ via $\phi_n(\zeta) = \sqrt{\frac{ \sqrt{|Q_f B|}}{2^n n! \sqrt{\pi}}} e^{-\frac{\zeta^2|Q_f B|}{2}} H_n\left(\zeta\sqrt{|Q_f B|}\right)$. We use throughout the paper the definition $x^\mu=(x_0,x_1,x_2,x_3)$ and $p^\mu=(p_0,p_1,p_2,p_3)$.

At mean field level, the quark mass $m_q=m_0-2G\langle\bar\psi\psi\rangle$ is controlled by the gap equation,
\begin{equation}
\label{gap}
m_q(1-2GJ_1)-m_0=0
\end{equation}
with
\begin{eqnarray}
J_1 &=& N_c\sum_{f,n}\alpha_n \frac{|Q_f B|}{2\pi} \int \frac{d p_3}{2\pi} \frac{1-F(E_f^+)-F(E_f^-)}{ E_f},
\end{eqnarray}
the summation over all flavors and Landau energy levels, spin factor $\alpha_n=2-\delta_{n0}$, quark energy $E_f=\sqrt{p^2_3+2 n |Q_f B|+m_q^2}$ and $E_f^\pm=E_f\pm \mu_f$, Fermi-Dirac distribution function $F(x)=\left( e^{x/T}+1 \right)^{-1}$, and the number of colors $N_c=3$ which is trivial in the NJL model. In our paper, the terminologies, chiral crossover or first order chiral phase transition are conventionally defined by the continuous change or the sudden jump of the order parameter $m_q$, and the connection point is the critical end point.

As quantum fluctuations above the mean field, mesons are constructed through quark bubble summation in the frame of random phase approximation (RPA)~\cite{njl2,njl1,njl3,njl4,njl5,zhuang}. Namely, the quark interaction via a meson exchange is effectively described by using the Dyson-Schwinger equation,
\begin{equation}
{\cal D}_M(x,z)  = 2G \delta(x-z)+\int d^4y\ 2G \Pi_M(x,y) {\cal D}_M(y,z),
\label{dsequ}
\end{equation}
where ${\cal D}_M(x,y)$ represents the meson propagator from $x$ to $y$, and the corresponding meson polarization function is the quark bubble,
\begin{equation}
\label{bubble}
\Pi_M(x,y) = i\ { {\text{Tr}}}\left[\Gamma_M^{\ast} S(x,y) \Gamma_M  S(y,x)\right]
\end{equation}
with the meson vertex
\begin{equation}
\label{vertex} \Gamma_M = \left\{\begin{array}{ll}
1 & M=\sigma\\
i\tau_+\gamma_5 & M=\pi^+ \\
i\tau_-\gamma_5 & M=\pi^- \\
i\tau_3\gamma_5 & M=\pi^0\ \ \ ,
\end{array}\right.\ \ \ \
\Gamma_M^* = \left\{\begin{array}{ll}
1 & M=\sigma\\
i\tau_-\gamma_5 & M=\pi^+ \\
i\tau_+\gamma_5 & M=\pi^- \\
i\tau_3\gamma_5 & M=\pi^0\ \ \ .
\end{array}\right.
\end{equation}
The quark propagator matrix $S=diag(S_u,\ S_d)$ in flavor space is at mean field level (see Eq.(\ref{quark})), and the trace is taken in spin, color and flavor spaces.

\subsection{Neutral mesons}
\label{sec:np}

When studying the chiral symmetry, we focus on its pseudo-Goldstone mode $\pi^0$ and Higgs mode $\sigma$. The neutral mesons $\pi^0$ and $\sigma$ are affected by external magnetic field only through the pair of charged constituent quarks. As a consequence, the transformation from coordinate space to momentum space is a conventional Fourier transformation, characterized by the plane wave $e^{-ik \cdot x}$~\cite{ritus1,ritus2,ritus21,beffect,beffect1,beffect2},
\begin{eqnarray}
\label{fourier1}
{\cal D}_{M}(k) &=&  \int d^4 (x-y) e^{i k \cdot (x-y)} {\cal D}_{M}(x,y),\nonumber\\
\Pi_{M}(k) &=&  \int d^4 (x-y) e^{i k \cdot (x-y)} \Pi_{M}(x,y),
\end{eqnarray}
for $M=\pi^0, \sigma$. By taking the quark bubble summation in RPA and considering the complete and orthogonal conditions of the plane wave $e^{-i k \cdot x}$, the neutral meson propagator in momentum space can be simplified as
\begin{equation}
\label{npole}
{\cal D}_M(k)=\frac{2G}{1-2G\Pi_M(k)}.
\end{equation}

The meson pole mass $m_M$ is defined as the pole of the propagator at zero momentum ${\bf k}={\bf 0}$,
\begin{equation}
\label{mmass}
1-2G\Pi_M(\omega^2=m^2_{M}, {\bf k}^2=0)=0,
\end{equation}
and the polarization function can be simplified
\begin{eqnarray}
\label{pi}
\Pi_M(\omega^2,0) = J_1-(\omega^2-\epsilon_M^2) J_2(\omega^2)
\end{eqnarray}
with $\epsilon_{\pi^0}=0$, $\epsilon_\sigma=2m_q$ and
\begin{eqnarray}
J_2(\omega^2) = -N_c\sum_{f,n}\alpha_n \frac{|Q_f B|}{2\pi} \int \frac{d p_3}{2\pi}{{1-F(E_f^+)-F(E_f^-)}\over  E_f (4 E_f^2-w^2)}.
\end{eqnarray}
During the chiral symmetry restoration, the quark mass decreases, and the $\pi^0$ mass keeps increase, as guaranteed by the Goldstone's theorem~\cite{gold1,gold2}. When the $\pi^0$ mass is beyond the threshold $m_{\pi^0} \geq 2m_q$, the decay channel $\pi^0 \rightarrow q {\bar q}$ opens, which defines the pion Mott transition~\cite{mott1,mott2,mott3}.

In chiral limit with vanishing current quark mass $m_0=0$, by comparing the gap equation (\ref{gap}) for quark mass $m_q$ with the pole equation (\ref{mmass}) for neutral meson pole mass $m_M$, we have the simple relation in the chiral symmetry breaking phase,
\begin{equation}
m_q \neq 0,\ \ \ \ \ m_{\pi^0} = 0,\ \ \ \ \ m_\sigma = 2m_q,
\end{equation}
and in the chiral restoration phase
\begin{equation}
m_q=0,\ \ \ \ \ m_{\pi^0} = m_\sigma \neq 0.
\end{equation}
This confirms that $\pi^0$ and $\sigma$ are the chiral partners in external magnetic field. $\pi^0$ is the Goldstone mode corresponding to the spontaneous chiral symmetry breaking, and $\sigma$ is the Higgs mode, which is heavier than (degenerate with) $\pi^0$ in chiral breaking (restoration) phase. When chiral restoration phase transition happens, the $\pi^0$ Mott transition happens simultaneously and the Higgs mode $\sigma$ approaches the minimum mass $m_\sigma|_{\min}=0$.

In physical case with non-vanishing current quark mass $m_0 \neq 0$, the chiral restoration is no longer a genuine phase transition. We observe (shown in Fig.\ref{figb20masscrossover} as an example) $m_q \gg m_0$, $m_{\pi^0}>0$ and $m_\sigma>2m_q$ in the region with spontaneous chiral symmetry breaking, $m_q > m_0$, $m_{\pi^0} \simeq 2m_q$ and $m_\sigma|_{\min}>0$ in the chiral restoration process, and $m_q \rightarrow m_0$ and $m_\sigma \simeq m_{\pi^0}\gg 2m_q$ in the region with partially restored chiral symmetry. Different definitions for the chiral restoration are proposed in the literatures~\cite{zhuang,phi,pipi,maocpc}, such as the maximum change of quark mass, the Mott transition of pseudo-Goldstone mode, and the minimum mass of Higgs mode. However, it should be mentioned that there is no guarantee for the coincidence of different definitions.

\subsection{Charged mesons}
\label{sec:cp}

For charged mesons $\pi^\pm$, we should take into account of the interaction between charged mesons and magnetic fields, which is absent for neutral mesons. With Ritus scheme, the Fourier transformation for neutral mesons (\ref{fourier1}) is extended to~\cite{coppola,maocharge,ritus1,ritus2,ritus21}
\begin{eqnarray}
\label{fourier2}
{\cal D}_M(\bar k) &=& \int d^4x d^4y F^\ast_k(x){\cal D}_M(x,y) F_k(y),\nonumber\\
\Pi_M(\bar k) &=& \int d^4x d^4y F^\ast_k(x)\Pi_M(x,y) F_k(y),
\label{ritubp}
\end{eqnarray}
where  ${\bar k}=(k_0,0,-s_M \sqrt{(2l+1)|Q_M B|},k_3)$ is the conserved four-dimensional Ritus momentum, and $F_k(x)=e^{-i \tilde k\cdot x} g^{s_M}_l(x_1,k_2)$ is the solution of the Klein-Gordon equation in magnetic field with the index $l$ describing meson Landau level, the Fourier transformed momentum $\tilde k=(k_0,0,k_2,k_3)$ and the meson sign factor $s_M=\text{{sgn}}(Q_M B)$.

Considering the complete and orthogonal conditions of $F_k(x)$ and the Dyson-Schwinger equation (\ref{dsequ}), the $\pi^\pm$ meson propagator in Ritus momentum space can be simplified as
\begin{equation}
{\cal D}_M(\bar k)=\frac{2G}{1-2G \Pi_M(\bar k)}.
\label{mb}
\end{equation}
The pole mass $m_M$ of charged pions is defined through the singularity of the meson propagator ${\cal D}^{-1}_M(\bar k)=0$ at $k_0=m_M, l=0$ and $k_3=0$,
\begin{equation}
\label{mpmass}
1-2G\Pi_M(m_M,0,-s_M\sqrt{|Q_M B|},0)=0.
\end{equation}

When considering the pion superfluid phase transition at $\mu_I>0$, we focus on the $\pi^+$ meson, because it acts as the Goldstone boson corresponding to the spontaneous breaking of isospin symmetry. Taking the mean field quark propagator (\ref{quark}) and the definition (\ref{bubble}) for the quark bubble, we have the $\pi^+$ polarization function at the pole,
\begin{eqnarray}
\label{pipm}
\Pi_{\pi^+}(k_0,0,-\sqrt{|eB|},0) = J_1+J_3(k^2_0)
\end{eqnarray}
with
\begin{eqnarray}
\label{eq8}
J_3(k^2_0) &=& \sum_{n,n'} \int \frac{d p_3}{2\pi}\frac{j_{n,n'}(k_0)}{4E_n E_{n'}}\\
&& \times \left[\frac{F(-E_{n'}-\mu_u)- F(E_n-\mu_d)}{k_0+\mu_I+E_{n'}+E_n}+\frac{F(E_{n'}-\mu_u)- F(-E_n-\mu_d)}{k_0+\mu_I-E_{n'}-E_n}\right],\nonumber\\
j_{n,n'}(k_0) &=& \left[{(k_0+\mu_I)^2/2}-n'|Q_u B|-n|Q_d B|\right]j^+_{n,n'} -2 \sqrt{n'|Q_u B|n|Q_d B|}\ j^-_{n,n'}~,\nonumber
\end{eqnarray}
and quark energy $E_{n'}=\sqrt{p^2_3+2 n' |Q_u B|+m_q^2}$ and $E_n=\sqrt{p^2_3+2 n |Q_d B|+m_q^2}$.

Pion superfluid phase transition at $\mu_I>0$ is a genuine phase transition with massless Goldstone mode $\pi^+$. At weak magnetic field and vanishing temperature and baryon chemical potential, by straightforwardly comparing the gap equation (\ref{gap}) of quark mass and pole equation (\ref{mpmass}) of $\pi^+$ mass, the critical isospin chemical potential for pion superfluid phase transition $\mu_I^\pi$ is equal to the $\pi^+$ mass in magnetic fields~\cite{maopion}.

\subsection{Pauli-Villars regularization}
Because of the four-fermion interaction, the NJL model is not a renormalizable theory and needs regularization~\cite{njl2,njl1,njl3,njl4,njl5}. The external magnetic field does not cause extra ultraviolet divergence but introduces discrete Landau levels and anisotropy in momentum space. To avoid the unphysical oscillations and the breaking of the law of causality under external magnetic field, we make use of the covariant Pauli-Villars regularization scheme~\cite{mao1,mao11,regula2}. In this scheme, the quark momentum runs formally from zero to infinity, and the divergence is removed by the cancellation among the subtraction terms. One introduces the regularized quark masses $m_i=\sqrt{m^2_q+a_i\Lambda^2}$ for $i=0,1,\cdots, N$, and replaces $m^2_q$ in the quark energy $E_f$ by $m_i^2$ and the summation and integration are changed as
\begin{eqnarray}
\sum_n\int \frac{dp_3}{2\pi} Function(E_f) \rightarrow \sum_n\int \frac{dp_3}{2\pi} \sum_{i=0}^N \left[ c_i \times Function(E_f^i) \right], \nonumber
\end{eqnarray}
where the coefficients $a_i$ and $c_i$ are determined by constraints $a_0=0$, $c_0=1$, and $\sum_{i=0}^N c_im_i^{2L}=0$ for $L=0,1,\cdots N-1$. The parameters used in our numerical calculations are $N=3$, $a_1=1, c_1=-3$, $a_2=2, c_2=3$, $a_3=3, c_3=-1$. The three parameters in the NJL model, namely the current quark mass $m_0=5$ MeV, the coupling constant $G=3.44$ GeV$^{-2}$ and the mass parameter $\Lambda=1.127$ GeV, are fixed by fitting the quark condensate $\langle\bar\psi\psi\rangle=-2 \times (250\ \text{MeV})^3$, pion mass $m_\pi=134\ \text{MeV} $ and pion decay constant $f_\pi=93\ \text{MeV}$ in vacuum~\cite{njl2,njl1,njl3,njl4,njl5}.

\section{Results and discussions}
\label{sec:rd}

\subsection{chiral partners $\pi^0$ and $\sigma$ in $T-\mu_B$ plane}

Chiral symmetry is spontaneously broken in vacuum, and will be restored at finite temperature and chemical potential. Since $\mu_B$ and $\mu_I$ play similar roles in chiral restoration, we will fix $\mu_I=0$ in this section. The chiral crossover happens in high temperature and low baryon chemical potential region, the first order chiral phase transition happens in low temperature and high baryon chemical potential region, and there exists a critical end point.

%%%%%%%%%%%%%%%%%%%%%%%%%%%%%%%%%%%%%%%%%%%%%%%%%%%%%%%%%%%%%%%%%%%%%%%%%%
\begin{figure*}[htb]
\begin{center}
\includegraphics[width=7.5cm]{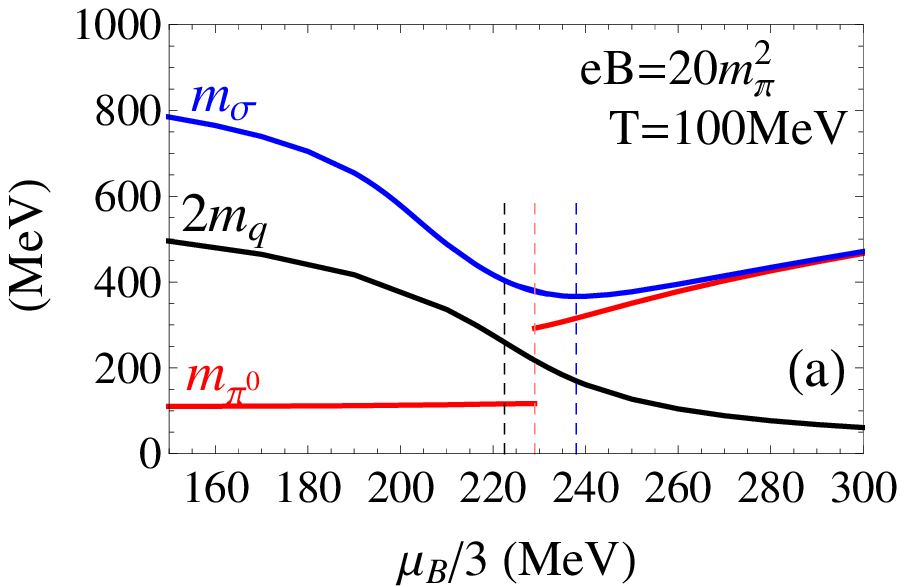}
\includegraphics[width=7.5cm]{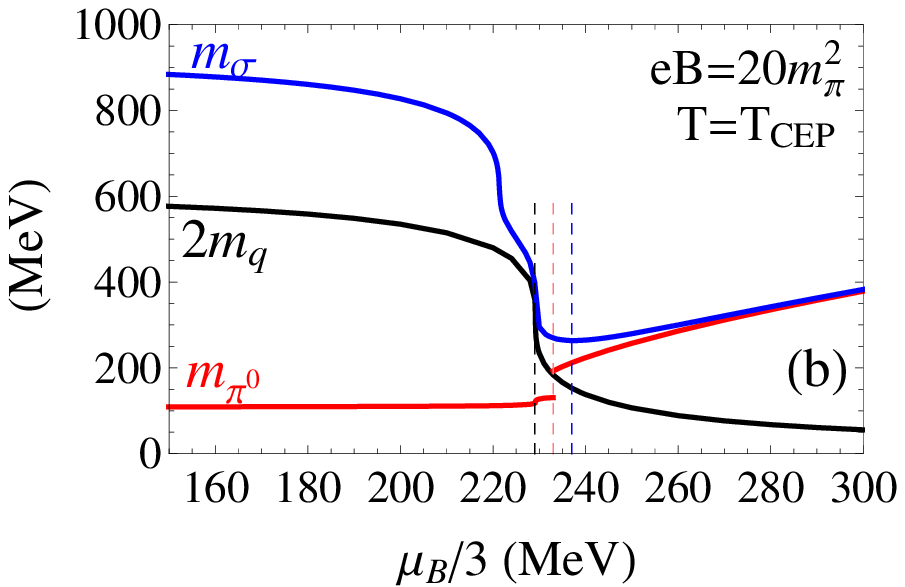}
\includegraphics[width=7.5cm]{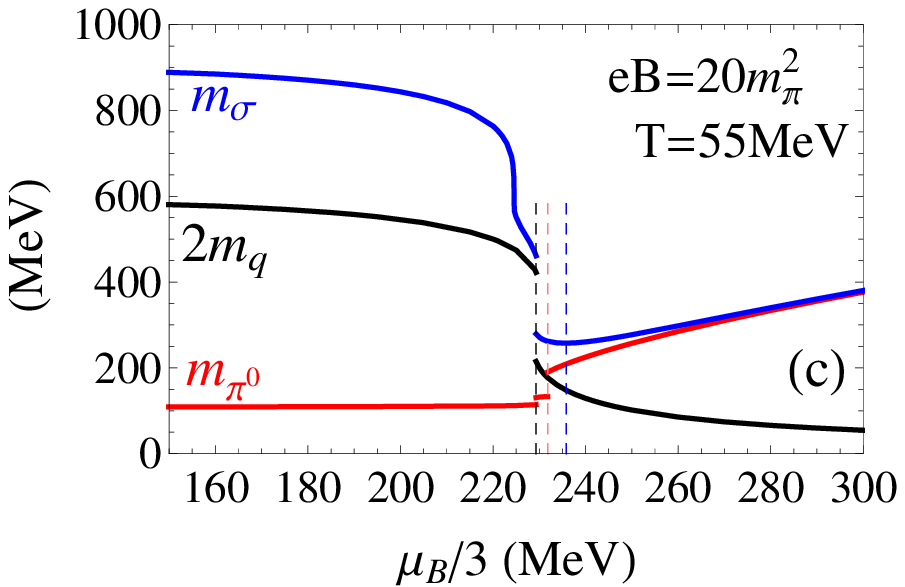}
\includegraphics[width=7.5cm]{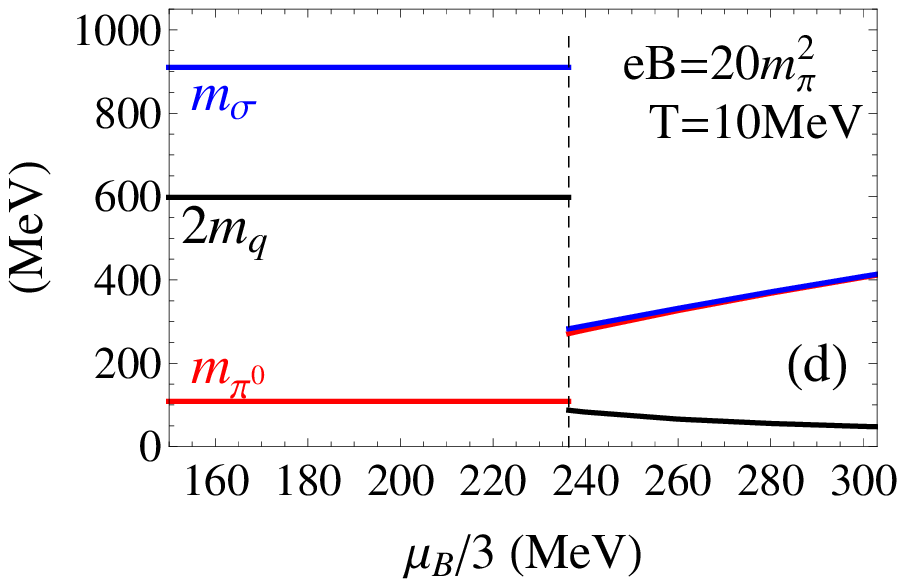}
\end{center}
 \caption{Masses of quark, $\pi^0$ and $\sigma$ with $eB=20m^2_\pi$ around the critical end point. Panel $(a)$ is an example in chiral crossover region with $T=100$ MeV, the panel $(b)$  at critical end point with $T=T_{\texttt{{CEP}}}=59$ MeV, and the panels $(c),(d)$  in first order chiral phase transition region with $T=55,\ 10$ MeV. The vertical dashed lines are used to denote the maximum change of quark mass (in black), the $\pi^0$ Mott transition (in red) and the minimum value of $\sigma$ mass (in blue).}
 \label{figb20masscrossover}
\end{figure*}
%%%%%%%%%%%%%%%%%%%%%%%%%%%%%%%%%%%%%%%%%%%%%%%%%%%%%%%%%%%%%%%%%%%%%%%%%%

Associated with the chiral crossover at zero baryon chemical potential, the $\pi^0$ meson has a sudden mass jump at the Mott transition induced by the discrete Landau level of constituent quarks, and with the first order chiral phase transition at zero temperature, the $\pi^0$ meson also shows a sudden mass jump induced by the mass jump of constituent quarks~\cite{mao2,q2,q3,yulang2010.05716}. What is the situation in the region with finite temperature and baryon chemical potential?  Fig.\ref{figb20masscrossover} plots the masses of quark, $\pi^0$ and $\sigma$ meson with $eB=20m^2_\pi$ around the critical end point. Here panel $(a)$ is an example in chiral crossover region with $T=100$ MeV, panel $(b)$  at critical end point with $T=T_{\texttt{{CEP}}}=59$ MeV, and panels $(c),(d)$ in first order chiral phase transition region with $T=55,\ 10$ MeV. In the chiral crossover region and at the critical end point, the quark and $\sigma$ masses continuously change with baryon chemical potential, but the $\pi^0$ meson has a sudden mass jump at the Mott transition. In the first order chiral phase transition region, the masses of quark, $\pi^0$ and $\sigma$ meson all have jumps. The vertical dashed lines in Fig.\ref{figb20masscrossover} are used to denote the maximum change of quark mass (in black), the $\pi^0$ mass jump at Mott transition (in red) and the minimum value of $\sigma$ mass (in blue).

In chiral crossover region, as shown by Fig.\ref{figb20masscrossover}$(a)$, we have smooth decrease of quark mass with the maximum change at $\mu_B/3=\mu^{q}_{\texttt{pc}}$. Due to the discrete quark Landau level in the magnetic field, there exists a mass jump for the pseudo-Goldstone mode $\pi^0$ at the Mott transition point $\mu_B/3=\mu^{\pi^0}_{\texttt{mott}}$, where the $\pi^0$ mass suddenly jumps up from $m_{\pi^0}<2m_q$ to $m_{\pi^0}>2m_q$. Except this jump, the $\pi^0$ mass keeps increasing during the chiral restoration process, which is consistent with the decreasing quark mass, as guaranteed by the Goldstone's theorem~\cite{gold1,gold2}. The Higgs mode $\sigma$ is always in the resonant state with $m_\sigma>2m_q$. $m_\sigma$ decreases in the chiral breaking phase, and reaches the minimum at $\mu_B/3=\mu^\sigma_{\texttt{min}}$. Then, it turns to increase and becomes almost degenerate with $\pi^0$ meson in the chiral restoration phase. We numerically obtain $\mu^{q}_{\texttt{pc}}<\mu^{\pi^0}_{\texttt{mott}}<\mu^\sigma_{\texttt{min}}$. In Fig.\ref{figb20masscrossover}$(b)$, we fix temperature at the critical end point $T_{\texttt{CEP}}$, and plot the masses of quark, $\pi^0$ and $\sigma$ as functions of baryon chemical potential. At critical end point $T=T_{\texttt{CEP}}$ and $\mu_B/3=\mu_{\texttt{CEP}}$, the quark mass, $\pi^0$ and $\sigma$ meson masses show the sharpest change, with $\frac{d m_q}{d \mu_B}\rightarrow \infty$, $\frac{d m_{\pi^0}}{d \mu_B}\rightarrow \infty$ and $\frac{d m_{\sigma}}{d \mu_B}\rightarrow \infty$, as indicated by the black dashed vertical line. Different from the continuous mass change of quark and $\sigma$ meson with baryon chemical potential, $\pi^0$ shows a sudden mass jump after the critical end point at $\mu_B/3=\mu^{\pi^0}_{\texttt{mott}}>\mu_{\texttt{CEP}}$, which is the Mott transition with mass jumping from $m_{\pi^0}<2m_q$ to $m_{\pi^0}>2m_q$, as denoted by the red dashed vertical line. The $\sigma$ meson, which is in the resonate state, reaches its minimum mass at a higher baryon chemical potential, with $\mu_B/3=\mu^\sigma_{\texttt{min}}>\mu^{\pi^0}_{\texttt{mott}}>\mu_{\texttt{CEP}}$, see the blue dashed vertical line.

Different from the single mass jump of $\pi^0$ meson in the chiral crossover region and at the critical end point, we observe twice $\pi^0$ mass jumps in the first order chiral phase transition region. As shown by Fig.\ref{figb20masscrossover}$(c)$, in the first order chiral phase transition region close to the critical end point, the quark mass has a sudden jump at $\mu_B/3=\mu^{q}_{\texttt{pc}}$, which leads to the mass jumps of $\pi^0$ and $\sigma$ meson, as indicated by the black dashed vertical line. After this mass jump, the $\pi^0$ meson is still in the bound state with $m_{\pi^0}<2m_q$. There appears a second mass jump for the $\pi^0$ meson at $\mu_B/3=\mu^{\pi^0}_{\texttt{mott}}>\mu^{q}_{\texttt{pc}}$, denoted by the red dashed vertical line, which is the Mott transition with the mass jumping from $m_{\pi^0}<2m_q$ to $m_{\pi^0}>2m_q$. For $\sigma$ meson, which is in the resonate state, after the mass jump induced by the quark mass jump, its mass keeps decreasing, approaches the minimum value at $\mu_B/3=\mu^\sigma_{\texttt{min}}>\mu^{q}_{\texttt{pc}}$, and then starts to increase with baryon chemical potential. Nevertheless, in the first order chiral phase transition region with very low temperature, see Fig.\ref{figb20masscrossover}$(d)$, associated with the quark mass jump, the $\pi^0$ Mott transition occurs and the $\sigma$ meson jumps to its minimum mass simultaneously, with $\mu^{q}_{\texttt{pc}}=\mu^{\pi^0}_{\texttt{mott}}=\mu^\sigma_{\texttt{min}}$. The quark mass jump at the first order chiral phase transition becomes larger with lower temperature, and the induced mass jumps for $\pi^0$ and $\sigma$ meson will also become larger. Comparing Fig.\ref{figb20masscrossover}$(c)$ and Fig.\ref{figb20masscrossover}$(d)$, when the constituent quark has large enough mass jump, the associated $\pi^0$ mass jump will satisfy the condition of Mott transition, jumping up from $m_{\pi^0}<2m_q$ to $m_{\pi^0}>2m_q$, and therefore, we only observe single $\pi^0$ mass jump in Fig.\ref{figb20masscrossover}$(d)$. And for $\sigma$ meson in the resonate state, the mass will jump down directly to its minimum value, and then start to increase, as shown in Fig.\ref{figb20masscrossover}$(d)$. In $T-\mu_B$ plane, the mass jump of pseudo-Goldstone mode $\pi^0$ can be induced either by the Mott transition (discrete Landau level of constituent quarks) or by the first order chiral phase transition (mass jump of constituent quarks), and the mass jump of Higgs mode $\sigma$ is only caused by the first order chiral phase transition.

%%%%%%%%%%%%%%%%%%%%%%%%%%%%%%%%%%%%%%%%%%%%%%%%%%%%%%%%%%%%%%%%%%%%%%%%%%
\begin{figure*}[htb]
\begin{center}
\includegraphics[width=10cm]{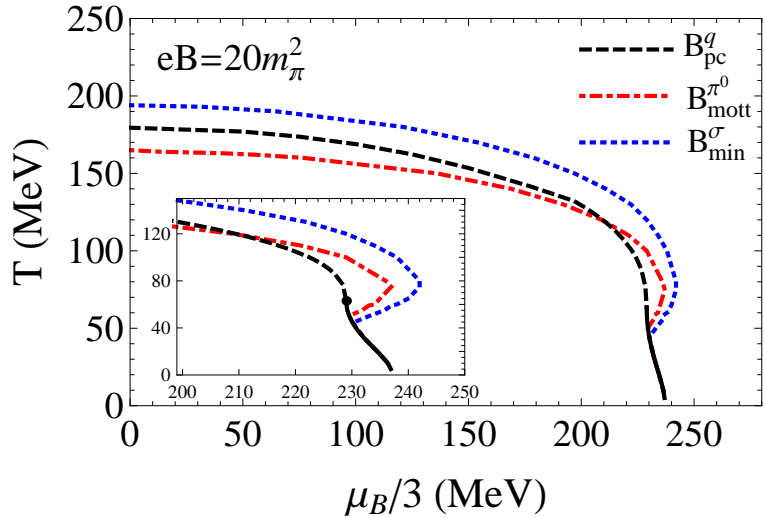}
\end{center}
 \caption{The chiral phase diagram in $T-\mu_B$ plane at finite external magnetic field $eB=20m^2_\pi$ with three characteristic phase boundaries,  $\texttt{B}^q_{\texttt{pc}}$ defined by the change of quark mass (black line), $\texttt{B}^{\pi^0}_{\texttt{mott}}$ by the Mott transition of pseudo-Goldstone boson $\pi^0$ (red dash-dotted line), and $\texttt{B}^\sigma_{\texttt{min}}$ by the minimum mass of Higgs mode $\sigma$ (blue dotted line). Here, the terminologies, chiral crossover (black dashed line), critical end point (black point) or first order chiral phase transition (black solid line) are conventionally defined from the continuous change or sudden jump of the order parameter $m_q$.}
 \label{figb20phasediagram}
\end{figure*}
%%%%%%%%%%%%%%%%%%%%%%%%%%%%%%%%%%%%%%%%%%%%%%%%%%%%%%%%%%%%%%%%%%%%%%%%%%

The order parameter of spontaneous chiral symmetry breaking is the quark mass $m_q$, and the corresponding pseudo-Goldstone mode and Higgs mode is $\pi^0$ and $\sigma$ meson, respectively. Therefore, we can, on the one hand, define chiral restoration from order parameter, and on the other hand, from the $\pi^0$ and $\sigma$ meson. We depict in Fig.\ref{figb20phasediagram} the chiral phase diagram in $T-\mu_B$ plane at $eB=20m^2_{\pi}$ with three characteristic phase boundaries $\texttt{B}^q_{\texttt{pc}}$, $\texttt{B}^{\pi^0}_{\texttt{mott}}$, $\texttt{B}^\sigma_{\texttt{min}}$. The phase boundary $\texttt{B}^q_{\texttt{pc}}$ is defined from the quark mass, and we denote the first order chiral phase transition by the black solid line and the chiral crossover by the black dashed line, with a critical end point located at $(T,\mu_B/3)=(T_{\texttt{CEP}},\mu_{\texttt{CEP}})=({59\ \text{MeV},229\ \text{MeV}})$. The phase boundaries $\texttt{B}^{\pi^0}_{\texttt{mott}}$ and $\texttt{B}^\sigma_{\texttt{min}}$ are defined by the Mott transition of pseudo-Goldstone boson $\pi^0$ and by the minimum mass of Higgs mode $\sigma$, respectively. They show apparent bump structure around the critical end point. In the chiral crossover region with high temperature and low baryon chemical potential, we have $\mu^{\pi^0}_{\texttt{mott}}<\mu^q_{\texttt{pc}}<\mu^\sigma_{\texttt{min}}$ for the characteristic baryon chemical potential on the phase boundaries when fixing temperature. As getting close to the critical end point from the crossover side, there appears a crossing for the two phase boundaries $\texttt{B}^q_{\texttt{pc}}$ and $\texttt{B}^{\pi^0}_{\texttt{mott}}$, and we thus obtain $\mu^q_{\texttt{pc}}<\mu^{\pi^0}_{\texttt{mott}}<\mu^\sigma_{\texttt{min}}$ for the characteristic baryon chemical potential with fixed temperature, which is also observed in the first order chiral phase transition region near the critical end point. In the first order chiral phase transition region with low temperature and high baryon chemical potential, the three phase boundaries become degenerate. It is noticeable that the starting point of the overlap between $\texttt{B}^q_{\texttt{pc}}$ and $\texttt{B}^{\pi^0}_{\texttt{mott}}$ is deviated from that for $\texttt{B}^q_{\texttt{pc}}$ and $\texttt{B}^\sigma_{\texttt{min}}$.% The chiral crossover happens in high temperature region with continuous change of quark mass, the first order chiral phase transition happens in high baryon chemical potential region with sudden jump of quark mass, and there exists a critical end point.

In the physical world with non-vanishing current quark mass, the chiral symmetry is an approximate symmetry, and hence its restoration is not a genuine phase transition. As shown in Fig.\ref{figb20phasediagram}, the phase boundaries from order parameter side and from meson side are different. In this case, one may obtain some results which are in conflict with the original idea of chiral symmetry restoration. For instance, the pseudo-Goldstone mode may become resonant state (or still be in bound state) in chiral breaking (or restoration) phase defined by the order parameter, which breaks down the Goldstone's theorem~\cite{gold1,gold2}. The critical end point characterized by the sharpest continuous change of quark mass and meson mass is different from the Mott transition of the pseudo-Goldstone mode. The location of the minimum $\sigma$ mass is deviated from the Mott transition of pseudo-Goldstone mode $\pi^0$ and the maximum change of quark mass. How to take a suitable definition for the phase boundary of chiral symmetry restoration in the $T-\mu_B$ plane is still an open question.

\subsection{Goldstone bosons $\pi^+$ and $\pi^0$ in $T-\mu_I$ plane}

This section focuses on the pion superfluid phase transition and chiral symmetry restoration in the $T-\mu_I$ plane under external magnetic field, which are determined by the corresponding Goldstone mode $\pi^+$ and pseudo-Goldstone mode $\pi^0$, respectively. Different from chiral symmetry, the isospin symmetry $U(1)_I$ is strict, and hence pion superfluid phase transition can be equivalently defined through the order parameter (charged pion condensate) and Goldstone mode (massless $\pi^+$ meson), as guaranteed by the Goldstone's theorem~\cite{gold1,gold2,lqcd2,model9}. In our current work, to avoid the complication and difficulty of dealing with charged pion condensate under external magnetic field, we start from the normal phase only with neutral chiral condensate, calculate the $\pi^+$ mass, and determine pion superfluid phase structure by the massless Goldstone boson $m_{\pi^+}=0$. For chiral symmetry restoration, since the characteristic phase boundaries from order parameter and pseudo-Goldstone boson are not far from each other, we plot the phase boundary defined through pseudo-Goldstone mode $\pi^0$, as parallel to the pion superfluid phase boundary defined by the Goldstone boson $\pi^+$.

%%%%%%%%%%%%%%%%%%%%%%%%%%%%%%%%%%%%%%%%%%%%%%%%%%%%%%%%%%%%%%%%%%%%%%%%%%
\begin{figure*}[htb]
\begin{center}
\includegraphics[width=8cm]{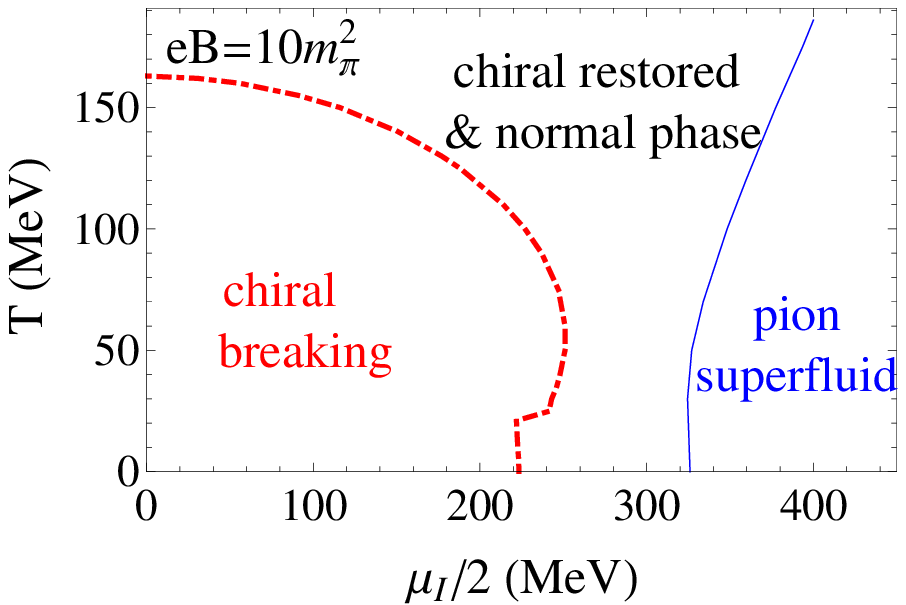}
\includegraphics[width=8cm]{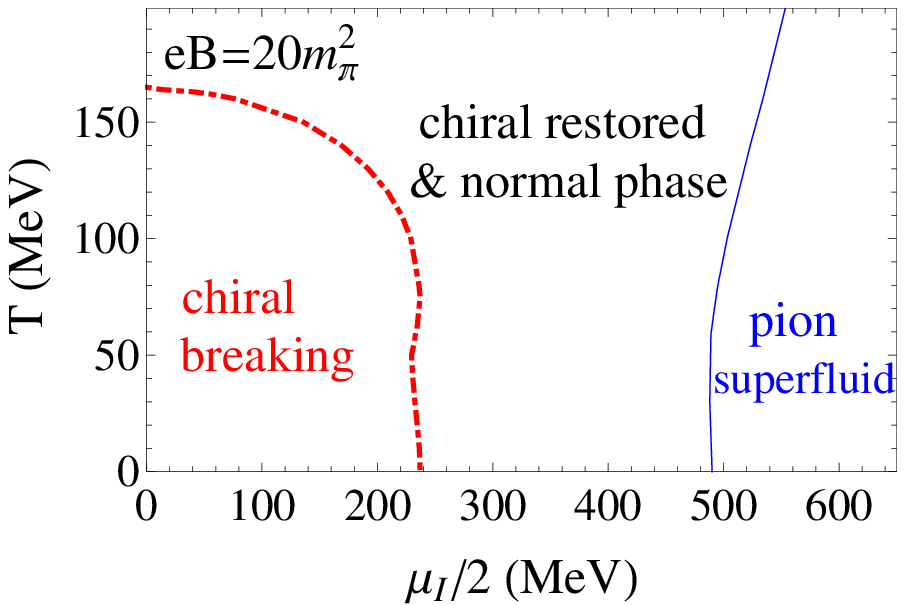}
\end{center}
 \caption{Phase diagram of pion superfluid and chiral restoration in $T-\mu_I$ plane at $eB=10m^2_\pi$ (upper panel) and $eB=20m^2_\pi$ (lower panel), where the pion superfluid phase transition line (blue solid line) is determined by the massless Goldstone boson $\pi^+$ ($m_{\pi^+}=0$), and the phase boundary of chiral symmetry restoration (red dash-dotted line) is determined by the Mott transition of the pseudo-Goldstone boson $\pi^0$ ($m_{\pi^0}=2m_q$). }
 \label{figbpionsuperfluid}
\end{figure*}
%%%%%%%%%%%%%%%%%%%%%%%%%%%%%%%%%%%%%%%%%%%%%%%%%%%%%%%%%%%%%%%%%%%%%%%%%%

The phase diagram in $T-\mu_I$ plane is depicted in Fig.\ref{figbpionsuperfluid} with $eB=10m^2_\pi$ (upper panel) and $eB=20m^2_\pi$ (lower panel). On the one hand, isospin chemical potential tends to break the isospin symmetry and restore the chiral symmetry. On the other hand, temperature tends to enhance the quark thermal motion, and leads to the phase transition from pion superfluid phase to normal phase and chiral symmetry restoration. In the low temperature and low isospin chemical potential region, the system is in the chiral breaking and normal phase, which means that the chiral symmetry is spontaneously broken and isospin symmetry is intact. In the low temperature and high isospin chemical potential region, the chiral symmetry is restored and the isospin symmetry is spontaneously broken. The system is in chiral restored and pion superfluid phase. In between with the medium isospin chemical potential, there is a chiral restored and normal phase. With high enough temperature, the system is in the chiral restored and normal phase, due to the strong quark thermal motion. Different from the apparent bump structure of chiral restoration phase boundary, the phase transition temperature of pion superfluid slightly decreases with isospin chemical potential in low temperature region, and then increases in high temperature region. As increasing magnetic field, the separation between the two phase boundaries becomes larger, that is the region of chiral restored and normal phase is enlarged. This is mainly caused by the fast increase of $\pi^+$ mass under external magnetic field, due to the electromagnetic interaction~\cite{maocharge,maopion}.%, where the pion superfluid phase transition line (blue solid line) is determined by the massless Goldstone boson $\pi^+$ ($m_{\pi^+}=0$), and the phase boundary of chiral symmetry restoration (red dash-dotted line) is determined by the Mott transition of the pseudo-Goldstone boson $\pi^0$ ($m_{\pi^0}=2m_q$)

%%%%%%%%%%%%%%%%%%%%%%%%%%%%%%%%%%%%%%%%%%%%%%%%%%%%%%%%%%%%%%%%%%%%%%%%%%
\begin{figure*}[htb]
\begin{center}
\includegraphics[width=7.5cm]{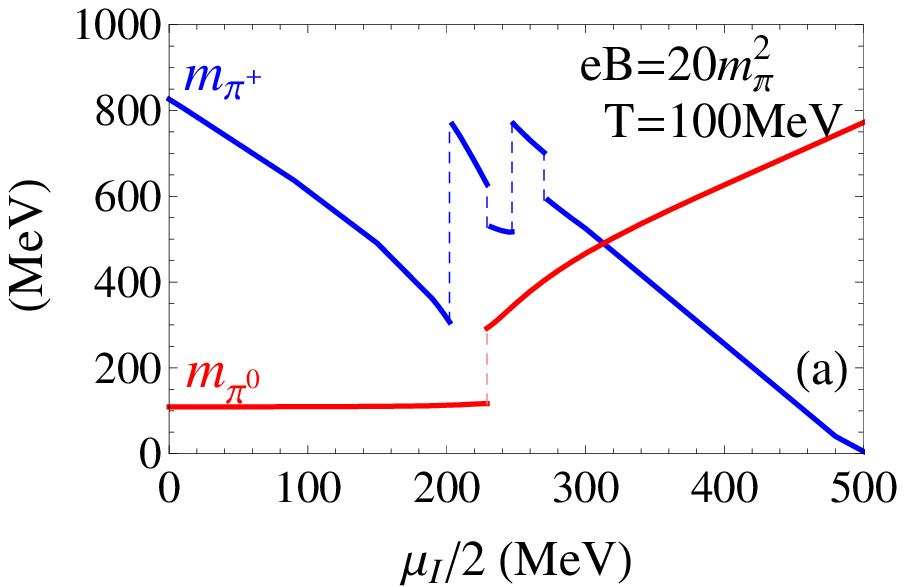}
\includegraphics[width=7.5cm]{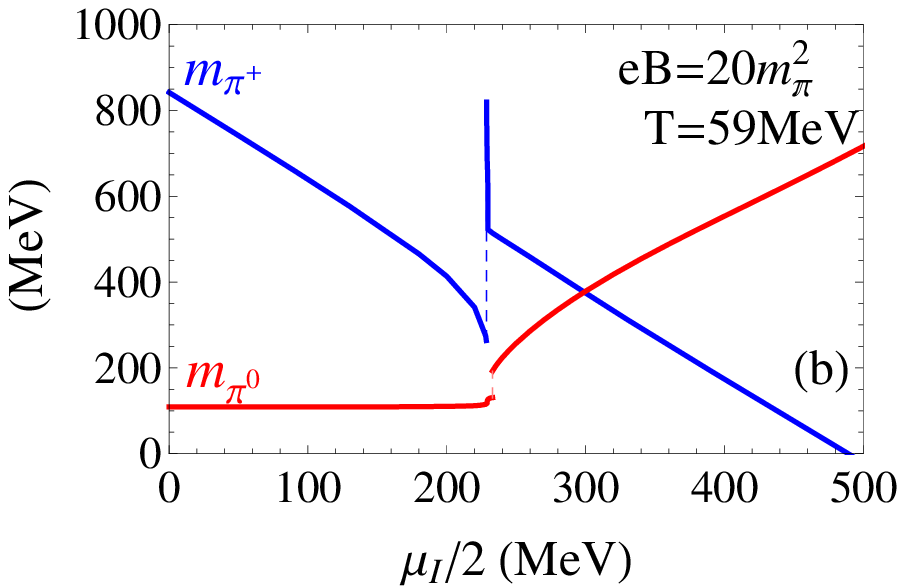}
\includegraphics[width=7.5cm]{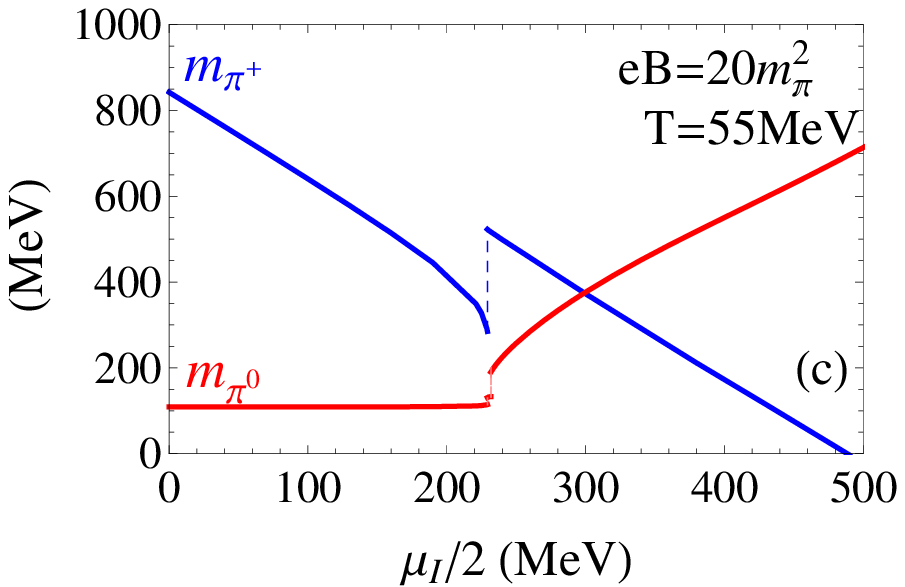}
\includegraphics[width=7.5cm]{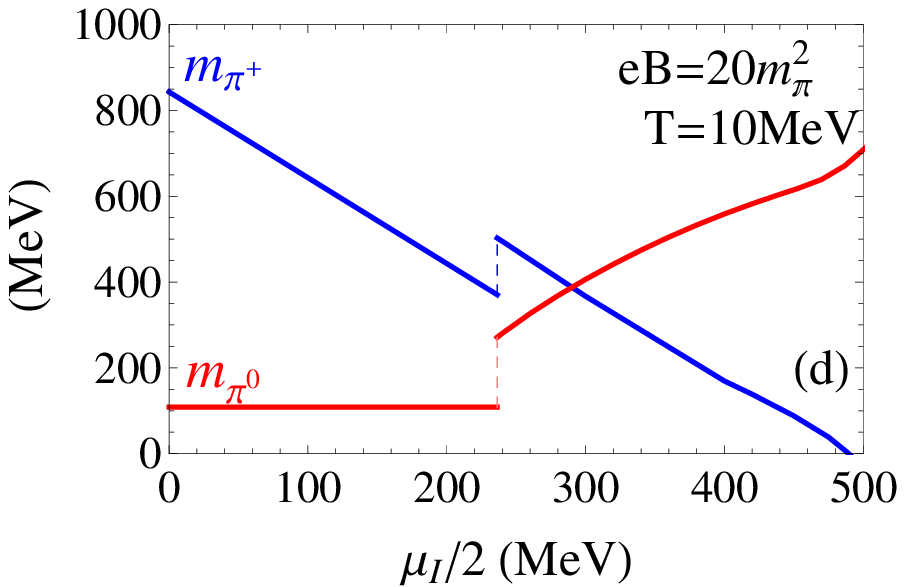}
\end{center}
 \caption{Masses of $\pi^+$ and $\pi^0$ with $eB=20m^2_\pi$ as functions of isospin chemical potential $\mu_I/2$, in chiral crossover region with $T=100$ MeV in panel $(a)$, at critical end point with $T=T_{\texttt{{CEP}}}=59$ MeV in panel $(b)$, and in first order chiral phase transition region with $T=55,\ 10$ MeV in panels $(c),(d)$. The vertical dashed lines are used to denote the $\pi^0$ mass jump (in red) and $\pi^+$ mass jump (in blue).}
 \label{figb20mpimui}
\end{figure*}
%%%%%%%%%%%%%%%%%%%%%%%%%%%%%%%%%%%%%%%%%%%%%%%%%%%%%%%%%%%%%%%%%%%%%%%%%%

Fig.\ref{figb20mpimui} makes comparison between the mass of $\pi^+$ (Goldstone boson of isospin symmetry breaking) and $\pi^0$ (pseudo-Goldstone boson of chiral symmetry breaking) in $T-\mu_I$ plane under external magnetic field. Here, we choose the same magnetic field $eB=20m^2_\pi$ and temperatures as shown in Fig.\ref{figb20masscrossover}, which represent the chiral crossover region with $T=100$ MeV in panel $(a)$, critical end point with $T=T_{\texttt{{CEP}}}=59$ MeV in panel $(b)$, and first order chiral phase transition with $T=55,\ 10$ MeV in panels $(c),(d)$. The isospin and baryon chemical potential play the same roles for chiral symmetry restoration, and thus the $\pi^0$ mass and quark mass are the same as in Fig.\ref{figb20masscrossover}, only with the replacement between $\mu_B/3$ and $\mu_I/2$. Due to the electromagnetic interaction between $\pi^+$ meson and external magnetic field, the $\pi^+$ mass becomes heavier than $\pi^0$ meson at zero isospin chemical potential. When increasing isospin chemical potential $\mu_I$, the isospin symmetry will be broken, which leads to the decrease of $\pi^+$ mass down to zero, but the broken chiral symmetry will be restored, which leads to the increase of $\pi^0$ mass. Therefore, it is expected to observe the crossing behavior of the $\pi^+$ and $\pi^0$ mass, and the $\pi^0$ mass will become heavier than $\pi^+$ meson with high enough isospin chemical potential. It is noted in Fig.\ref{figb20mpimui} that independent of temperature, both $\pi^+$ and $\pi^0$ mass show approximate linear behavior as functions of isospin chemical potential in low and high $\mu_I$ regions.

In the medium $\mu_I$ region, there appear mass jumps for both $\pi^+$ and $\pi^0$, which depend on the temperature. These mass jumps are caused either by the discrete Landau levels of constituent quarks or by the quark mass jump. For example, with $T=100$ MeV in Fig.\ref{figb20mpimui}$(a)$, where the quark mass continuously decreases with isospin chemical potential, $\pi^+$ meson shows four times mass jumps induced by the discrete quark Landau levels, and this is different from the single mass jump of $\pi^0$ meson. Due to different electric charges of constituent quarks, the $\pi^+$ meson may have several mass jumps depending on the temperature and isospin chemical potential, which are determined by condition $m_{\pi^+}+\mu_I=\sqrt{2n'|Q_u|B+m_q}+\sqrt{2n|Q_d|B+m_q}$ with $n' \geq 0,\ n\geq 1$, according to Eq.(\ref{pipm}). And for $\pi^0$ meson, the mass jump occurs with condition $m_{\pi^0}=2m_q$. In Fig.\ref{figb20mpimui}$(b)$ with $T=T_{\texttt{{CEP}}}=59$ MeV, the $\pi^+$ mass jump occurs a little bit earlier than the critical end point $\mu_I^{\texttt{CEP}}$, and it is also earlier than the mass jump of $\pi^0$. After the mass jump, there is a very sharp mass decreasing of $\pi^+$ meson (see the almost vertical blue solid line), which is induced by the sharp change of constituent quark mass around $\mu_I^{\texttt{CEP}}$. Finally, it approaches a slower and constant mass decrease rate. It should be mentioned that the sequence of $\pi^+$ mass jump and the critical end point $\mu_I^{\texttt{CEP}}$ is dependent on the magnetic field. We have numerically checked that at $eB=10m^2_\pi$, the $\pi^+$ mass jump happens later than its corresponding $\mu_I^{\texttt{CEP}}$. In the first order chiral phase transition region near critical end point with $T=55$ MeV in Fig.\ref{figb20mpimui}$(c)$, we obtain single $\pi^+$ mass jump caused by the quark mass jump, which is different from the twice mass jumps of $\pi^0$ meson induced by the quark mass jump and the discrete quark Landau levels, respectively. With lower temperature $T=10$ MeV in Fig.\ref{figb20mpimui}$(d)$, both $\pi^+$ and $\pi^0$ mesons have only single mass jump due to the constituent quark mass jump at first order chiral phase transition. In $T-\mu_I$ plane, different from the twice mass jumps of $\pi^0$ in the first order chiral phase transition region, the $\pi^+$ meson displays several times mass jumps in the chiral crossover region. At the critical end point, they both show very sharp but continuous mass changes. %As a consequence of such jumps, it may result in some interesting phenomena in relativistic heavy ion collisions. For instance, when the formed fireball cools down, there might be sudden enhancement of pions. How to extract the experimental signal for the critical end point should be further investigated in the future.

\section{Summary}
\label{sec:sum}
%Chiral symmetry restoration and pion superfluid phase transition are investigated in magnetized NJL model. Different from the conventional study of order parameters, we focus on the meson properties in $T-\mu_B-\mu_I-eB$ space.

Light mesons $(\sigma, \pi^0, \pi^\pm)$ are investigated in $T-\mu_B-\mu_I-eB$ space by using a two-flavor NJL model, which are related to the chiral symmetry restoration and pion superfluid phase transition.

In $T-\mu_B$ plane, we plot the chiral phase diagram under external magnetic field by three characteristic phase boundaries, $\texttt{B}^q_{\texttt{pc}}$ defined by the maximum change of quark mass, $\texttt{B}^{\pi^0}_{\texttt{mott}}$ by the Mott transition of pseudo-Goldstone boson $\pi^0$, and $\texttt{B}^\sigma_{\texttt{min}}$ by the minimum mass of Higgs mode $\sigma$. On the phase boundary $\texttt{B}^q_{\texttt{pc}}$, we can distinguish chiral crossover, critical end point or first order chiral phase transition by the continuous change or the sudden jump of quark mass. During the chiral restoration process, the pseudo-Goldstone mode $\pi^0$ has the sudden mass jump. This is caused not only by the discrete quark Landau level, which is the Mott transition, but also by the sudden mass jump of constituent quarks, which is associated with the first order chiral phase transition. In the chiral crossover region, $\pi^0$ meson shows single mass jump induced by the discrete quark Landau level. In the first order chiral phase transition region with very low temperature, the single mass jump of $\pi^0$ meson is caused by the quark mass jump. At the critical end point, $\pi^0$ meson has a very sharp but continuous mass increase, together with a sudden mass jump at the Mott transition, and in the first order chiral phase transition region nearby, we observe twice $\pi^0$ mass jumps, induced by the discrete quark Landau level and quark mass jump, respectively. The Higgs mode $\sigma$, which is in the resonate state, has a non-monotonical mass change with a local minimum value. The mass is continuously changed in the chiral crossover region and at the critical end point, and shows a jump in the first order chiral phase transition region. Because of the explicit breaking of chiral symmetry in physical world, the phase boundaries from the order parameter side and from meson side are different from each other. It is still an open question to take a self-consistent definition for the chiral symmetry restoration. %The Mott transition of pseudo-Goldstone boson $\pi^0$ corresponds to its sudden mass jump under external magnetic field.

In $T-\mu_I$ plane, the competition between pion superfluid phase transition and chiral symmetry restoration is studied in terms of the corresponding Goldstone mode $\pi^+$ and pseudo-Goldstone mode $\pi^0$. The isospin symmetry is strict, and the pion superfluid phase transition is uniquely determined by the massless Goldstone mode $\pi^+$. On the one hand, isospin chemical potential tends to break the isospin symmetry and restore the chiral symmetry. On the other hand, temperature tends to induce the phase transition from pion superfluid phase to normal phase and chiral symmetry restoration. For the low $T$ case, the system is in the chiral breaking and normal phase with low $\mu_I$, in chiral restored and normal phase with medium $\mu_I$, and in chiral restored and pion superfluid with high $\mu_I$. The separation between the chiral restoration phase boundary and the pion superfluid phase transition is enhanced by the external magnetic field, due to the smooth decrease of $\pi^0$ mass and fast increase of $\pi^+$ mass. In $T-\mu_I$ plane, $\pi^+$ meson also has the sudden mass jump, caused either by the discrete quark Landau level or by the mass jump of constituent quarks. Different from the twice mass jumps of $\pi^0$ in the first order chiral phase transition region, several mass jumps of $\pi^+$ meson happen in the chiral crossover region. At the critical end point, they both show very sharp but continuous mass changes.

As a consequence of such mass jumps, it may result in some interesting phenomena in relativistic heavy ion collisions where strong magnetic field can be created, for instance, the sudden enhancement of pions. Moreover, how to extract the experimental signal for the critical end point will be investigated in the future.%For the high $T$ region, the system is in chiral restored and normal phase.

\section*{Acknowledgement}

The work is supported by the NSFC Grant 11775165.

%\bibliographystyle{JHEP}
%\bibliography{F:/FuChunE/myPhysics/articles/chunefu/JebRef_DataBase/References}
%name of.bib

\end{document}